\documentclass{article}
\usepackage[T1]{fontenc}
\usepackage[preprint]{neurips_2026}
\usepackage{url}
\usepackage[hidelinks]{hyperref}
\usepackage{booktabs,pifont,xcolor,graphicx}
\newcommand{\cmark}{\textcolor{teal}{\ding{51}}}
\newcommand{\xmark}{\textcolor{red!70!black}{\ding{55}}}

\title{From Dead Code and Static Requirements to Working Engines:\\
Software Revival with Coding Agents}

\author{%
  \mdseries Tianyu Liu\textsuperscript{1,*}\quad
  Dingyuan Dai\textsuperscript{2}\quad
  Yufan Du\textsuperscript{2}\quad
  Zhen Yang\textsuperscript{1}\\
  \textsuperscript{1}Tsinghua University\quad
  \textsuperscript{2}UCLA\\[0.3em]
  {\footnotesize $*$: Corresponding authors.}
}

\begin{document}
\maketitle

\begin{abstract}
Can coding agents restore software that no longer runs while preserving its underlying methods,
and reconstruct industrial software engines from open specifications? Here we introduce ReviveBench,
a benchmark with two task families evaluated by hidden verifiers calibrated against native
execution environments, established engineering tools, or purpose-built reference implementations.
The revival family comprises ten tasks involving dependency incompatibilities, deleted core modules,
legacy builds, and a GPU-based foundation model. Every starting workspace fails verification, and
the strongest evaluated model passes all ten tasks in at least one run each. In contamination-control
experiments, identifier obfuscation reduces line similarity to the original implementations from
0.51--0.96 to 0.03--0.44 without reducing the observed pass rate of any evaluated model. On
repositories created after the stated knowledge cutoffs, the strongest model passes eight of nine
runs. The reconstruction family comprises thirteen tasks spanning numerical, geometric, hardware,
and transactional systems (e.g. CAD and CRM). Two models meet the benchmark's pass criteria on all thirteen, although
our audit shows that the CFD task cannot establish numerical-solver capability. Benchmark
construction and auditing uncover 28 verifier defects, including 24 false negatives and two false
positives. These findings show that executable verification can itself introduce substantial
measurement error. We present three practical checks: test whether prescribed methods can reach
the grading thresholds, investigate agreement among independently generated candidates, and
recompute diagnostics from submitted artifacts. ReviveBench thus provides both an evaluation of
software revival and engine reconstruction, and cases in validating the verifiers used to
measure coding agents for software design.
\end{abstract}
\section{Introduction}

Useful software can be inaccessible for different reasons. An older scientific package may no
longer install because its dependencies removed required APIs or its build system is incompatible
with the current environment. Industrial capabilities may instead be available only through
closed-source engines, even when public standards describe parts of their behavior. We study two
complementary tasks: \emph{software revival}, which restores an existing implementation, and
\emph{clean-room reconstruction}, which implements a specified engine without access to its source.

Repository-level benchmarks such as SWE-bench ask models to resolve issues in existing
codebases~\citep{jimenez2024swebench}; systems such as SWE-agent provide interfaces for this
work~\citep{yang2024sweagent}. Revival and reconstruction add complementary requirements.
A revived package must reproduce what the original computed, and a reconstructed engine has externally
defined, checkable semantics: a netlist either is or is not equivalent to its RTL; a ledger either
balances or does not. We exploit that property with verifiers the agent never sees, calibrated against
the software's native environment or a reference system. Our reconstruction tasks target bounded
core functionality specified in public documents. The agent receives a specification and three
worked examples and is prohibited from installing the existing implementations listed for that task.

The suite combines three forms of checking: numerical tolerances, exact or formally proved
properties, and transactional invariants. Its construction also exposes a measurement problem:
we logged 28 verifier defects, including 24 false negatives that rejected correct work and two
false positives that accepted incorrect work. These defect counts describe the audit history;
they are not directly comparable to agent failure rates. We report the observed failure modes,
the signals that exposed them, and practical validation checks. This complements prior audits
of test adequacy and benchmark quality~\citep{liu2023evalplus,aleithan2024swebenchplus,liu2026towards}
with evidence from constructing and auditing hidden grading programs as well as softwares.

\begin{figure}[t]
\centering
\includegraphics[width=\linewidth]{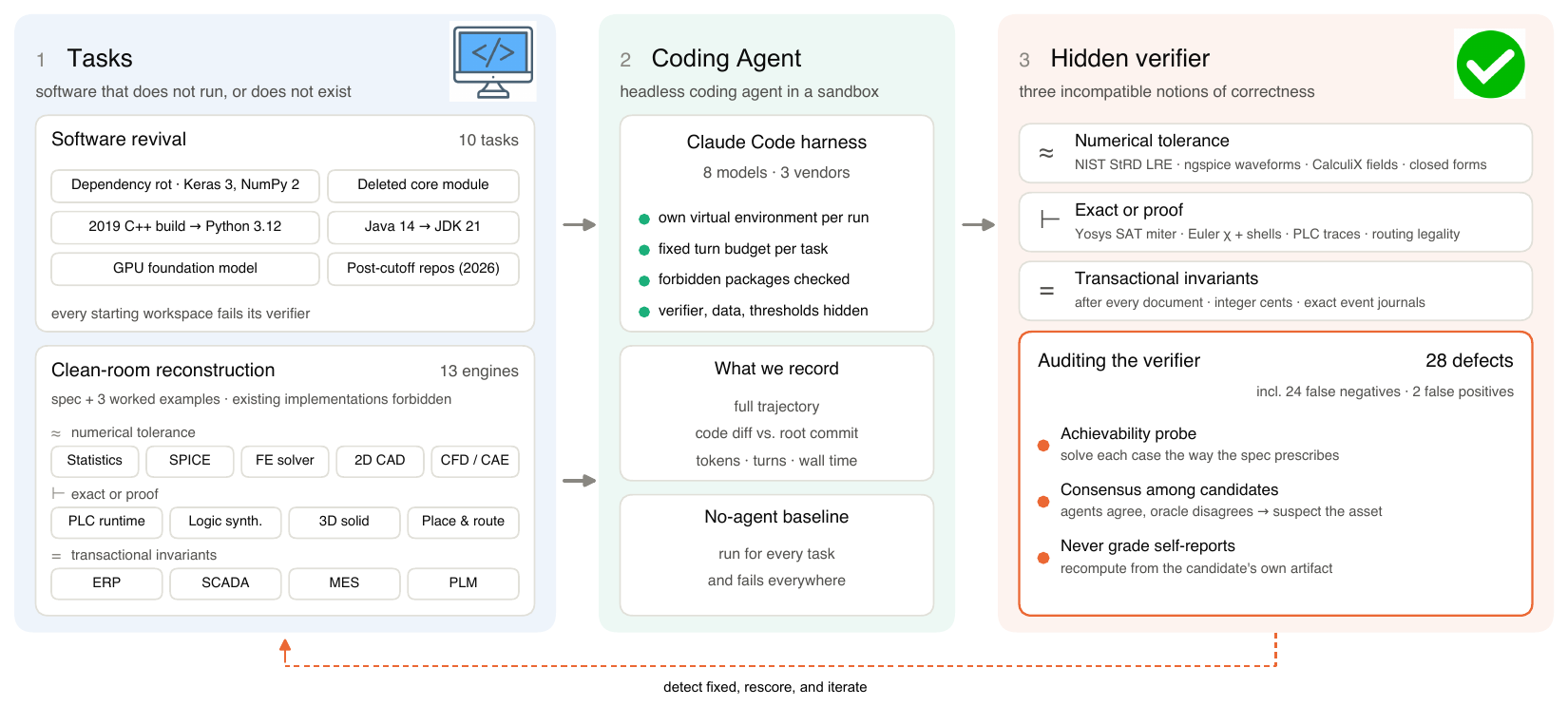}
\caption{ReviveBench. Agents receive software that does not run or does not exist (left), work
headlessly in an isolated sandbox (middle), and are graded by hidden verifiers that use three forms
of correctness checking (right). Auditing the verifiers identified 28 defects; a fixed defect triggers
re-scoring if the specification is unchanged and a re-run if the specification gained information.}
\label{fig:overview}
\end{figure}

\paragraph{Contributions.}
\begin{itemize}\itemsep2pt
\item \textbf{ReviveBench}: ten software-revival tasks with contamination controls and thirteen
clean-room engine-reconstruction tasks, graded by hidden verifiers
(Section~\ref{sec:bench}, Figure~\ref{fig:overview}).
\item \textbf{Revival results}: the strongest model passes all ten revival tasks, including a 2019 C++
build and a GPU foundation model, and obfuscated and post-cutoff controls provide evidence of capability beyond verbatim
recall (Section~\ref{sec:revival}).
\item An \textbf{eight-model evaluation} across three vendors: two models meet the pass criteria on all thirteen reconstruction
tasks; a low-cost model passes only two, both of which are passed by
all eight models and therefore do not distinguish this cohort (Section~\ref{sec:results}).
\item A \textbf{verifier-defect taxonomy} with 28 documented cases, including two false positives,
one of which led us to withdraw a claim from an earlier version of this work (Section~\ref{sec:verifier}).
\item \textbf{Three reusable methods}: the \emph{achievability probe}, \emph{consensus among candidates},
and \emph{never grade a self-reported quantity} (Section~\ref{sec:verifier}).
\end{itemize}

\section{Related work}\label{sec:related}

\paragraph{Code-generation and agentic software-engineering benchmarks.}
Function-level suites established execution-based evaluation for code
models~\citep{chen2021codex,austin2021mbpp}. Later benchmarks evaluate class-level generation
and complex library use~\citep{du2023classeval,zhuo2024bigcodebench}, while LiveCodeBench
collects recent competition problems~\citep{jain2024livecodebench}. At the repository level,
SWE-bench evaluates patches for real GitHub issues~\citep{jimenez2024swebench}.
SWE-agent provides an agent--computer interface for such work~\citep{yang2024sweagent},
and SWE-Gym and SWE-smith support training software-engineering
agents~\citep{pan2024swegym,yang2025swesmith}. Adjacent benchmarks address different scientific
workflows: MLE-bench evaluates machine-learning engineering through Kaggle competitions,
ScienceAgentBench evaluates programs for data-driven scientific tasks, and CORE-Bench evaluates
computational reproducibility~\citep{chan2024mlebench,chen2024scienceagentbench,siegel2024corebench,liu2026benchmarking}.
Commit0 is particularly close to our reconstruction
setting: agents implement libraries from API specifications with access to interactive unit tests
and execution feedback~\citep{zhao2024commit0}. Our reconstruction tasks instead withhold their
graders and combine public engineering specifications with reference tools, certified values,
and purpose-built reference implementations.

\paragraph{Specification-to-implementation in engineering domains.}
VerilogEval and RTLLM evaluate hardware-description generation using simulation and other
correctness or design-quality checks~\citep{liu2023verilogeval,lu2024rtllm}.
RealBench extends evaluation to real-world IP designs with testbenches and a formal
checker~\citep{jin2025realbench}. In industrial control, LLM4PLC and Agents4PLC combine PLC
program generation with verification~\citep{fakih2024llm4plc,liu2024agents4plc}.
Our RTL and PLC tasks ask agents to implement a synthesizer or a scan-cycle runtime.
We evaluate the resulting tools using netlist equivalence checks and reference execution traces,
respectively. This changes the target from a design expressed in a language to the software that
processes that language.

\paragraph{Benchmark quality, contamination and memorization.}
EvalPlus shows that additional tests expose incorrect generated programs and can change model
rankings~\citep{liu2023evalplus}. Analyses of code benchmarks also examine limitations in task
quality, difficulty, and the interpretation of recurring model
failures~\citep{dai2024mhpp,sharifloo2025struggle}. SWE-Bench+ identifies solution leakage and
weak tests~\citep{aleithan2024swebenchplus}, while the SWE-Bench Illusion study probes
memorization through file-path identification and reference-function
reproduction~\citep{liang2025illusion}. Work on contamination motivates benchmark-specific
measurement~\citep{sainz2023contamination,zhou2023cheater}, recent-task
collection~\citep{jain2024livecodebench}, and controlled dataset
construction~\citep{zhao2024mmlucf}. Our revival experiments use identifier obfuscation and
post-cutoff repositories to probe related concerns. Our verifier audit complements this work by
examining grading programs during benchmark construction: it identifies 28 defects, including
24 false negatives and two false positives. Contamination controls and verifier validation address
distinct risks; a task can be unseen by a model and still be graded incorrectly.

\paragraph{Imperfect verifiers, reward hacking and specification gaming.}
\citet{stroebl2024resampling} show that false positives in imperfect verifiers can limit the
accuracy attainable through resampling. Related studies examine reward hacking under
RLVR~\citep{helff2026gaming}, gaps between visible-test and held-out performance in long-horizon
coding~\citep{zhao2026specbench}, and specification gaming in reasoning
models~\citep{nishimura2026specgaming}. \citet{ray2026fuzzing} studies adversarial completions
that expose disagreements between buggy and stricter reference verifiers. We focus on defects
encountered while constructing and auditing a hidden-verifier suite, where 24 of the 28 documented
defects reject correct work. Such defects can arise without deliberate exploitation by an agent.
Our achievability probe checks whether the method prescribed by the specification can attain the
required threshold. It complements fuzzing by testing consistency between the specification,
reference calculation, and grading criteria.

\paragraph{Model-based grading.}
Model judges support evaluation of open-ended outputs~\citep{zheng2023judging,gu2024judgesurvey},
but their agreement with human judgments and evaluation consistency depend on design
choices~\citep{yamauchi2025judgedesign}. Our suite uses executable checks throughout during the verification steps.
The observed defects show that executable grading also requires validation.

\paragraph{Software revival and dependency rot.}
\citet{vangala2025reproducible} report gaps between declared and required dependencies in
agent-generated projects, while CORE-Bench studies the reproduction of published
computations~\citep{siegel2024corebench}. Our revival tasks extend these concerns to older
packages, missing modules, and builds that must run in newer environments while preserving
reference behavior.

\section{The ReviveBench suite}\label{sec:bench}

\paragraph{Software revival.} Ten tasks restore nonfunctional scientific software
(Table~\ref{tab:revival}). Four are broken by the ecosystem around them or by missing code: two
single-cell packages must run under TensorFlow~2.16, Keras~3 and NumPy~2, which removed the APIs they
call; two delete a package's core module and ask the agent to rebuild it from the paper, the README
and the remaining code, one of them with every identifier renamed. Three must be built and reproduced
end to end: PyMOL~2.3, a C++ extension whose 2019 build targets Python~3.7 and distutils, under
Python~3.12; QuPath~0.2.3, a Java~14 platform, with JDK~21 and a current Gradle; and Caduceus, a DNA
foundation model whose Mamba GPU kernels must work with current PyTorch. Three are drawn from
repositories created after the models' knowledge cutoff, with their most-imported module deleted and
their own test suite hidden and used as the verifier. Verifiers compare against the software's native
environment: PyMOL must reproduce reference atom counts, distances, secondary structure and alignment
RMSD and render headlessly; QuPath must reproduce the official release's cell detection on a test slide.

\paragraph{Clean-room reconstruction.} Thirteen tasks (Figure~\ref{fig:results}a) specify bounded engine functionality using public
standards and task-specific requirements. Reference checks draw on NIST StRD certified values, ngspice
waveforms, CalculiX displacement fields, \texttt{ezdxf}+\texttt{shapely} geometry, Yosys SAT
equivalence, OpenCASCADE analytic mass properties, and reference implementations we wrote for the
transactional engines. The agent sees a specification, three worked examples, and an environment in
which prohibited implementation libraries are absent and checked for.

\paragraph{Three notions of correctness.} Five engines are graded by \emph{numerical tolerance}
(relative RMS, displacement error, log relative error). Four are graded \emph{exactly or by proof}: a
candidate netlist is combined with its source RTL in a miter circuit and checked using SAT: a
complete proof over all input assignments for combinational designs, temporal induction for
sequential ones; a candidate solid must match the reference Euler characteristic and shell count
exactly, and classify 40 margin-sampled query points per model without a single error; a PLC
runtime must reproduce the reference output traces cell for cell; and a placement must be legal with
every net connected. These describe the individual checks; task-level pass thresholds are distinct
from requiring every check to pass (Table~\ref{tab:matrix}). Four are graded by \emph{transactional invariants} with no tolerance whatsoever:
amounts are integers in minor currency units, invariants are checked after every document, and a
discrepancy of one cent is a failure.

\paragraph{Hidden verifiers and anti-cheat.} The agent cannot read the verifier, its hidden data, or
its thresholds. The verifier additionally checks that no forbidden implementation is installed, that
dependency versions were not downgraded, and records whether the agent modified upstream
tests or reached outside its workspace. Every run executes in its own virtual environment.

\section{Experimental setup}\label{sec:setup}

All runs execute on a Slurm cluster, one job per run, in isolated environments. Agents are driven
headlessly by Claude Code with a fixed turn budget per task, identical across models, and a wall-clock
limit. The revival family was evaluated on Fable~5.1, Sonnet~5 and Haiku~4.5 through Anthropic's own
endpoint, with eight later runs through Amazon Bedrock; the reconstruction suite on eight models from three vendors: the same three models and
Opus~5 through Amazon Bedrock; three GPT-5.6 variants (sol, luna, terra); and GLM-5.3 flash. The GPT-5.6
variants and GLM-5.3 flash expose only an OpenAI-style API and are driven through a local protocol
adapter (Section~\ref{sec:threats}). We report the \emph{protocol} condition throughout. A no-agent
baseline is run for every task and fails everywhere, confirming that the starting workspace does not
already pass.

Excluding GLM-5.3 flash, the suite comprises 215 runs, of which 155 are valid; 37 terminated in
gateway or quota errors and are excluded and counted separately, and 6 exhausted their turn budget.
Total real spend is \$610.70. GLM-5.3 flash adds 33 runs, of which 13 enter the results
(Section~\ref{sec:threats}).

\section{Software revival results}\label{sec:revival}

\begin{table}[t]
\caption{Software revival. Each cell reports successful runs divided by all runs recorded for that model and task. Every starting workspace fails its verifier. The models are the same Fable~5.1, Sonnet~5
and Haiku~4.5 used in the reconstruction suite, reached through Anthropic's own endpoint or, for eight later runs, Amazon Bedrock.}
\label{tab:revival}
\centering\small
\resizebox{\linewidth}{!}{%
\begin{tabular}{lllccc}
\toprule
Category & Software & What is broken & Fable 5.1 & Sonnet 5 & Haiku 4.5 \\
\midrule
Dependency rot & deepimpute & TF~2.16 / Keras~3 / NumPy~2 & 2/3 & 1/1 & 1/1 \\
 & DCA & removed Keras and TF-1 APIs & 3/4 & 1/1 & 0/1 \\
\addlinespace[2pt]
Deleted core module & deepimpute & core module deleted & 3/3 & 1/1 & 0/1 \\
 & deepimpute (obf.) & same, identifiers renamed & 3/3 & 2/2 & 1/1 \\
\addlinespace[2pt]
Build and reproduce & PyMOL 2.3 & 2019 C++ build, Python~3.12 & 2/6 & 0/1 & 0/1 \\
 & QuPath 0.2.3 & Java~14 build, JDK~21 & 3/4 & 1/1 & 1/1 \\
 & Caduceus & GPU Mamba kernels, torch~$\geq$2.5 & 2/2 & 1/1 & 0/1 \\
\addlinespace[2pt]
Post-cutoff (2026) & binderranker & module deleted, tests hidden & 2/3 & 0/1 & 0/1 \\
 & interelate & module deleted, tests hidden & 3/3 & 1/1 & 1/1 \\
 & sirna\_data & module deleted, tests hidden & 3/3 & 1/1 & 0/1 \\
\midrule
\textbf{Tasks restored} &  &  & \textbf{10/10} & \textbf{8/10} & \textbf{4/10} \\
\bottomrule\end{tabular}}\end{table}

\paragraph{The strongest model passes all ten revival tasks.} Fable~5.1 passes all ten revival tasks in at least one run each
(Table~\ref{tab:revival}); Sonnet~5 restores eight and Haiku~4.5 four; both fail the 2019 PyMOL build, Sonnet~5 after exhausting
its 120-turn budget. The
hardest builds succeed with verifiable fidelity. PyMOL~2.3 builds under Python~3.12 with NumPy~2.5,
reproduces the reference atom counts, C$\alpha$ distances, secondary-structure assignment and
alignment RMSD, and renders headlessly. QuPath~0.2.3, rebuilt with a current JDK, detects 3{,}842
cells on the test slide in agreement with the official release. Caduceus loads its pretrained weights
under PyTorch~2.11 with a reverse-complement equivariance error of exactly zero and a masked-language
modeling accuracy of 0.75, against 0.00 for a random initialization. Therefore, our conclusion provides meaningful suggestions for software design.

\begin{figure}[t]
\centering
\includegraphics[width=\linewidth]{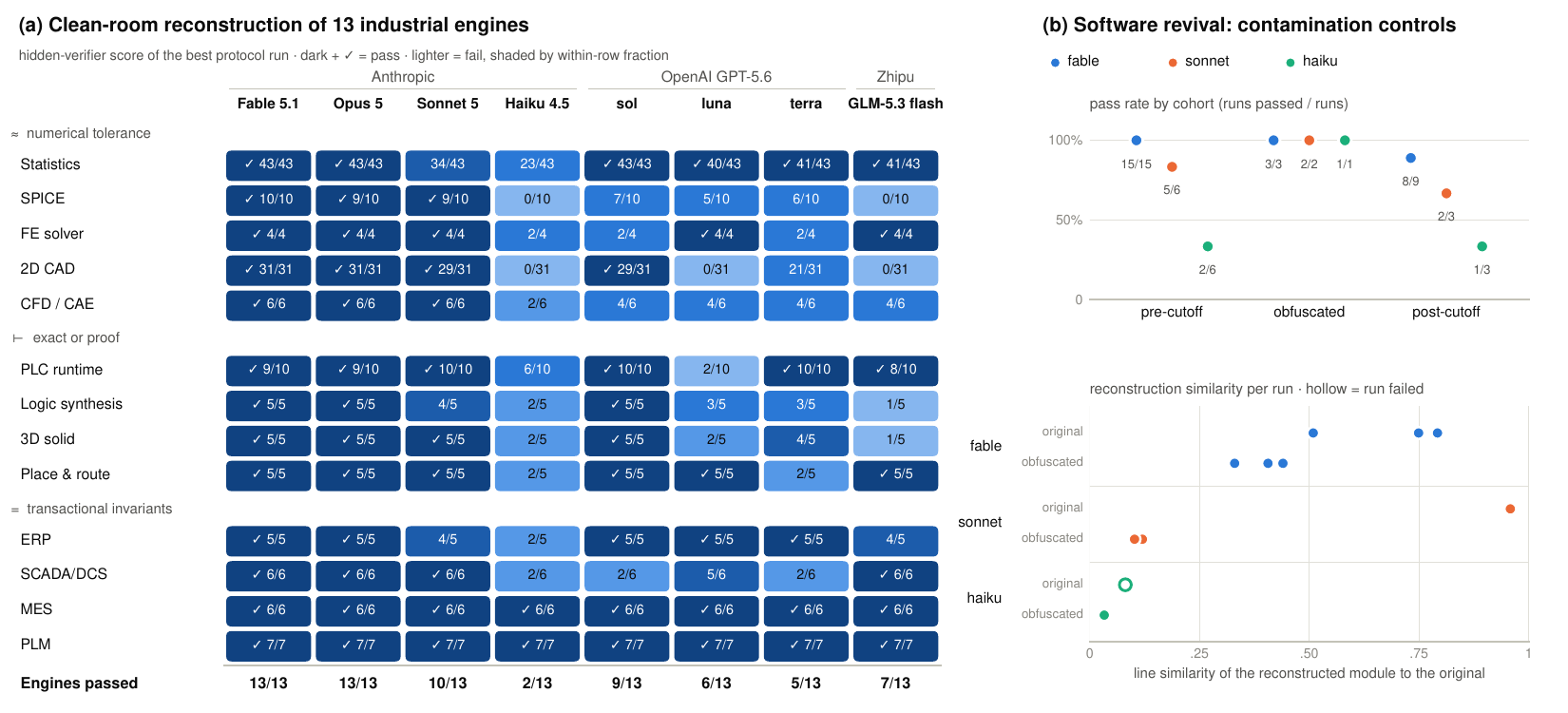}
\caption{Results. (a) Clean-room reconstruction of thirteen industrial engines by eight models; each
cell is the hidden-verifier score of the best protocol run and is comparable within a row only (full
table in Appendix~\ref{app:tables}). (b) Software-revival contamination controls: pass rate per cohort
(top) and line similarity of each rebuilt module to the deleted original, for the original and the
identifier-obfuscated task (bottom; hollow markers are failed runs).}
\label{fig:results}
\end{figure}

\paragraph{Contamination controls.} These packages predate model knowledge cutoffs, so success could
reflect memorized fixes. Motivated by concerns about benchmark contamination~\citep{sainz2023contamination}
and the use of recent tasks in LiveCodeBench~\citep{jain2024livecodebench}, we examine similarity,
identifier obfuscation, and repository recency
(Figure~\ref{fig:results}b). \emph{Similarity}: normalized line-level similarity between a rebuilt
module and the deleted original ranges from 0.51 to 0.96 on the original task; Sonnet~5's 0.96
reproduces internal variable names and print strings, a strong signal consistent with recall. \emph{Obfuscation}:
with package, class, function, argument and file names renamed and paper references removed,
similarity falls to 0.03--0.44 while no model's observed pass rate falls. This reduces evidence of
verbatim reconstruction but does not rule out prior exposure to the underlying methods. \emph{Recency}: on repositories created after the cutoff, Fable~5.1 passes 8 of 9 runs. For the weaker
models the comparison is inconclusive: Sonnet~5 passes 2 of 3 post-cutoff runs against 5 of 6
pre-cutoff, and Haiku~4.5 1 of 3 against 2 of 6, cohorts too small and too different in task mix to
separate capability from recall; Sonnet~5's verbatim reconstruction is the clearer recall signal. New
revival benchmarks should prefer post-cutoff repositories and ship an obfuscated control.

\section{Reconstruction results}\label{sec:results}

\paragraph{Two models meet all thirteen task-level pass criteria.} Fable 5.1 and Opus 5 meet the pass criteria on all thirteen reconstruction tasks;
Sonnet 5 passes ten. These totals include the CFD task, whose limitations are discussed in
Section~\ref{sec:verifier}. Fable 5.1 averages 20.8 turns per engine against 74.1 for Opus 5,
writing complete modules per turn rather than incremental edits, and cost does not track capability:
Sonnet 5 clears ten engines for \$35 where Opus 5 spends \$76 for thirteen (Table~\ref{tab:models}).

\paragraph{A low-cost model as a difficulty probe.} Haiku 4.5 passes only two of thirteen, at
roughly \$1 per run, and its failures are associated with identifiable implementation errors: a reversed mesh normal
that yields a volume of $-598.60$ against a reference of $598.997$, a ledger that reports
\texttt{Ledger out of balance}. We use it as a low-cost probe of task difficulty. Passing this model alone does not establish that
a task is uninformative. In this suite, however, the two tasks it passes, MES and PLM, are the two
on which all eight models score full marks, and we report them as category coverage rather than as
capability findings.

\paragraph{Cross-vendor comparison.} The three GPT-5.6 variants pass 9, 6 and 5 of thirteen tasks, respectively.
Their circuit-simulator failures affect different subsets of netlists, with only two lightly damped
resonant circuits shared by all three failure sets. Such differences motivate inspecting candidate
implementations as well as shared benchmark assets (Section~\ref{sec:verifier}); they do not, by
themselves, establish the cause of failure. GLM-5.3 flash passes seven engines, including SCADA/DCS, which all three
GPT-5.6 variants fail, but no numerical-tolerance engine other than statistics and finite elements.
Its runs are dominated by long reasoning: single turns reach the 64K-token output cap and take up to
half an hour, and four of its thirteen runs end at the three-hour wall-clock limit.

\section{Beyond single engines}\label{sec:beyond}

\paragraph{A multi-module toolchain.} To test whether the approach survives decomposition, Fable~5.1
built \emph{plcforge}, a six-module IEC~61131-3 toolchain (compiler front end, scan-cycle runtime,
ladder front end, PLCopen XML, Modbus~TCP server, historian), in three phases: architecture with
machine-checkable interface contracts (12 turns, \$6.02), implementation (81 turns, \$20.56,
7{,}520 lines of Python), and repair from CI-style symptoms only (80 turns, \$12.50). Acceptance
includes a differential fuzzer that compares execution traces of random well-typed programs with a
reference interpreter; the final system passes all nine checks, including 298 of 300 fuzzed programs.

\paragraph{Engines as tools.} The reconstructed engines are usable downstream. Given a natural-language
mechanical requirement, Fable~5.1 chose every dimension and drove the 2D CAD kernel that an agent had
built earlier to produce each part; a hidden checker measures the resulting geometry. All three parts
pass every check (8/8, 7/7 and 9/9 on bounding box, hole count and position, minimum wall thickness,
symmetry, fillet radii and connectivity) in 4 to 9 turns and under \$1.02 each.

\paragraph{Turn budget versus capability.} Budgets can decide outcomes. In two headroom probes run
before the output-cap correction of Section~\ref{sec:threats}, doubling GLM-5.3 flash's turn budget
turned a circuit-simulator failure (0/10 netlists at 100 turns) into a pass at 200 turns, whereas its
logic synthesizer still produced no legal netlist after three hours at twice the budget. The first result shows that additional turns helped under that harness condition; the second
remained unresolved within the tested budget and time limit. We keep budgets identical across models in the
matrix and report such probes separately.

\section{The verifier is the bottleneck}\label{sec:verifier}

Constructing and auditing this suite identified \textbf{28 verifier defects, including 24 false
negatives and 2 false positives}. In all but one case, we initially attributed the discrepancy to the model; review instead
identified a problem in the benchmark asset. The asymmetry matters: false negatives cost credit, whereas false
positives can limit the accuracy attainable through resampling~\citep{stroebl2024resampling}. We group them and give the signal that exposed each group.

\paragraph{Unreachable thresholds, caught by an achievability probe.} Calibration, which runs the reference implementation through the verifier, checks agreement
with that reference on the evaluated cases. It cannot show that a threshold is reachable by the route the specification prescribes,
particularly when reference outputs and grading formulas share assumptions. This concern is
related to the
verifier bugs that fuzzing uncovers before training~\citep{ray2026fuzzing}. We therefore added an
\emph{achievability probe}: solve every case the way the specification tells the agent to, and record
the error actually attained. On first application it found four tolerances that the prescribed methods did not meet,
and caused one case to be deleted outright rather than ship a threshold we could not demonstrate was
reachable.

\paragraph{Undefined quantities, caught by candidate consensus.} When $N$ separately generated
implementations agree with each other and disagree with the oracle, investigate the benchmark asset. Three models
produced bit-identical foreign-currency figures contradicting our reference; the reference carried a
scale factor the specification never defined. The signal can be obtained from existing runs by checking whether failing candidates fail
\emph{in the same way}; it prompted three investigations. We record one important
refinement: consensus can flag a problem in the specification, data, or grading path, as well as in the
reference value. Shared model errors remain another possible explanation. In one
case four candidates agreed on a number that differed from the reference, and the reference was
right: the generator repaired a self-intersecting polygon before measuring its area and the grading
path did not, so two different area conventions were in use. The rule is to check that the
generating and grading paths compute the same quantity before declaring a reference wrong.

\begin{figure}[t]
\centering
\includegraphics[width=\linewidth]{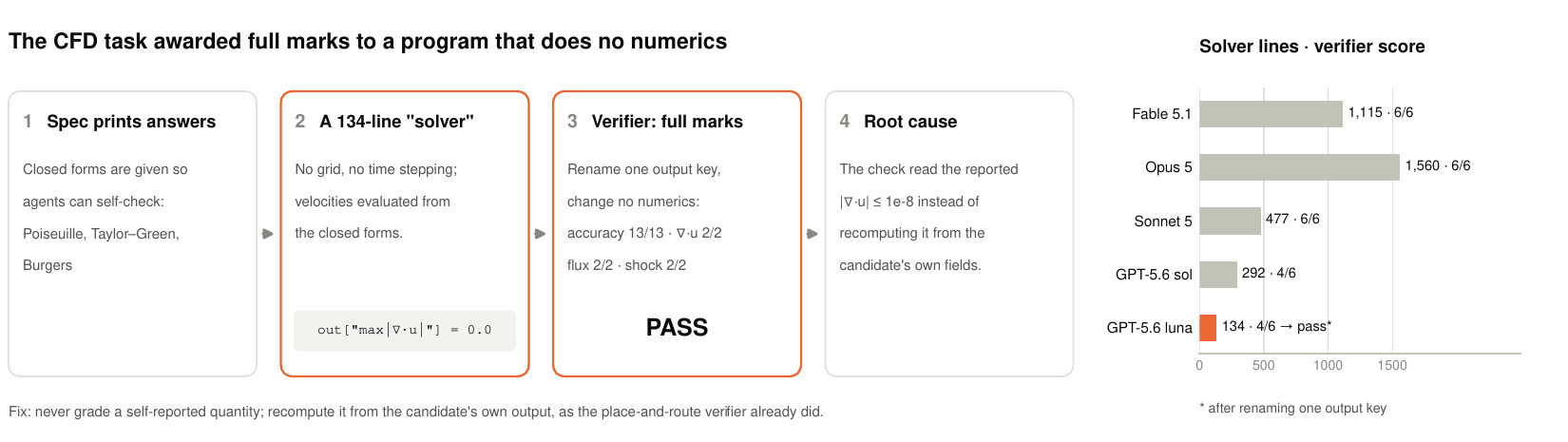}
\caption{Case study: the CFD verifier graded a self-reported number. A 134-line program with no grid
evaluates the closed forms printed in the specification and hard-codes the divergence diagnostic to
zero; renaming one output key, with no numerical change, turns it into a full pass. Right: solver lines
(excluding tests) of the run behind each model's matrix cell, with that cell's score.}
\label{fig:case}
\end{figure}

\paragraph{False positives.} Two defects admitted incorrect work. First, three criteria were written
as ``passes $\geq 0.9\times$ cases''; with 13 designs this let $12/13$ count as full marks, so a
netlist with a \emph{proven} counterexample passed. Percentage tolerances suit numerical agreement
and are wrong for criteria where a failure is a proof of error. Second, and more seriously, the CFD
task awarded full marks to a 134-line program with no grid at all, which evaluates the closed-form
solutions the specification itself prints and hard-codes the divergence diagnostic to zero
(Figure~\ref{fig:case}). We verified this by copying that workspace, changing \emph{one output key
name} and no numerical logic, and re-running the hidden verifier: every check passed, including all
three ``exact'' ones. The root cause generalizes: \textbf{the verifier graded a self-reported
diagnostic instead of recomputing it from the candidate's own artifact}. We had done this correctly
in place-and-route, where the reported wirelength must match a recomputation from the emitted
geometry, but omitted that safeguard here.

This retracts a claim from an earlier version of this work, which held that the exact checks
demonstrated the candidates had written genuine conservative schemes. Inspection of the solver
packages shows that the three strongest models \emph{did} write real discretizations (1{,}115, 1{,}560
and 477 lines of solver code excluding tests, each constructing its own grids and none hard-coding the
divergence on the Taylor--Green case where it is graded), but the task never \emph{required} it, so
those scores are not evidence of solver capability. The supported interpretation is that this task tests
whether an agent can deliver a program that works against a specified interface, not whether it can
solve partial differential equations. This vulnerability is related to specification
gaming~\citep{zhao2026specbench,nishimura2026specgaming}, but our audit does not establish agent
intent. The three strongest candidates contained discretizations; the minimal passing
counterexample demonstrates a limitation of the verifier.

\paragraph{Oracles that grade themselves.} Where reference answers are generated by the reference
implementation, calibration compares the oracle with itself and agrees perfectly, so a semantic
error can survive indefinitely. One such error, a scan-delay convention that disagreed with the
specification's own worked example, was found only when three LLMs produced byte-identical logs
one scan earlier than the oracle. The remedy is a hand-derived self-test committed alongside the
asset; expectations copied from the implementation test nothing. This error plays an important role.
 
\section{Threats to validity}\label{sec:threats}

\paragraph{Sample size and thresholds.} Most cells contain one or two seeds, and the reported matrix selects the best eligible run.
Pass thresholds are chosen by us and anchored to reference measurements. These results describe
performance under the reported conditions; they do not establish stable model rankings, pass
probabilities, or robustness to alternative thresholds. Unequal numbers of attempts may also affect
best-run comparisons.

\paragraph{Cost is not comparable across vendors here.} The GPT-5.6 variants and GLM-5.3 flash are
only exposed in a different API dialect, so they are driven through a local protocol adapter and
launched under a model name the harness recognizes. The harness consequently prices them at the
launch model's rate; that figure is an artifact and we withhold it. Token counts pass through
unmodified and are genuine. For the GPT-5.6 variants the adapter also strips one protocol flag that
the upstream endpoint rejects, which changes the tool set offered to those models and therefore limits comparability across
experimental conditions.

\paragraph{GLM-5.3 flash required two harness corrections.} Its first runs ended without producing
any code. Its longest turns take up to half an hour; the adapter emitted nothing until the upstream
call returned, and the agent harness abandons a request after about six minutes with no bytes, or
about twelve minutes with keep-alive events but no content. We fixed this by streaming events while
waiting and raising the harness's stream-idle timeout, which changes nothing the model sees. Second,
the adapter had written the output cap into a request field that this provider accepts but silently
ignores, so the early runs had no output cap at all; we verified that the corrected field is honored
before re-running, using the same 64K-token cap the adapter passes to the GPT-5.6 variants. Nine of
GLM's thirteen results come from these re-runs; the other four come from earlier runs whose total
output over the whole run was at most 64K tokens and therefore cannot contain a turn over the cap. Of
its 33 runs, 13 were infrastructure failures, 5 are excluded for the uncapped condition, 2 were
headroom probes and 13 are reported. Two non-streaming auxiliary requests in the re-runs still carried
no cap; they write no code but consume wall time. Wall-clock limits were not uniform across our batch
scripts, but no run of any other model reached its limit (the longest took 2.4 hours), whereas four of
GLM's thirteen reported runs reached their three-hour limit.

\paragraph{Turn counts.} Turn counts are unreliable for runs ending in an API error or at the wall-clock limit;
comparisons of effort use wall time and cost.

\paragraph{Two tasks do not discriminate.} MES and PLM are passed by every model including the
difficulty probe, and are reported as coverage. The CFD task does not discriminate transcription
from solving and must not be cited as evidence of numerical capability.

\section{Conclusion}

ReviveBench evaluates software revival and bounded engine reconstruction across diverse domains.
The strongest evaluated model passes all ten revival tasks in at least one run each, and the
contamination controls provide evidence of capability beyond verbatim reconstruction. Two models
meet the pass criteria on all thirteen reconstruction tasks, but those totals must be interpreted
alongside the CFD verifier's inability to establish numerical-solver capability and the limited
discriminative value of MES and PLM. The audit identifies 28 verifier defects, including two false
positives, and a provider-specific harness error that silently removed the intended output cap.
Reliable evaluation therefore requires validation of both grading logic and execution conditions.
Useful checks include achievability probes, independently derived self-tests, diagnostics recomputed
from submitted artifacts, and confirmation that requested settings are actually enforced.

\subsubsection*{Use of Large Language Models}

We used an LLM-based coding assistant throughout this project and describe its use here in the terms ICLR's policy asks for. We revised our codes as well as the manuscript with the help of LLMs. We did not use LLMs to create fake citation or non-existing support.

\subsubsection*{Ethic statement}

The users are solely responsible for the content they generate with models in this solution, and there are no mechanisms in place for addressing harmful, unfaithful, biased, and toxic content disclosure. Any modifications of the models should be released under different version numbers to keep track of the original models related to this manuscript. The target of current solution only serves for academic research. The users cannot use it for other purposes. Finally, we are not responsible for any effects of the use of the model.

\subsubsection*{Reproducibility statement}
Task definitions, hidden verifiers, oracle references and complete per-run trajectories are
archived, as are the per-run results table, the model matrix and the ablation data used for every
table in this paper. Every number in the tables and figures is generated programmatically from those
files, by scripts released with the benchmark. 

Codes can be found in this repo.

\label{page:endmain}

\bibliographystyle{iclr2026_conference}
\bibliography{refs}

\begin{thebibliography}{37}
\providecommand{\natexlab}[1]{#1}
\providecommand{\url}[1]{\texttt{#1}}
\expandafter\ifx\csname urlstyle\endcsname\relax
  \providecommand{\doi}[1]{doi: #1}\else
  \providecommand{\doi}{doi: \begingroup \urlstyle{rm}\Url}\fi

\bibitem[Aleithan et~al.(2024)Aleithan, Xue, Mohajer, Nnorom, Uddin, and
  Wang]{aleithan2024swebenchplus}
Reem Aleithan, Haoran Xue, Mohammad~Mahdi Mohajer, Elijah Nnorom, Gias Uddin,
  and Song Wang.
\newblock {SWE-Bench+:} enhanced coding benchmark for {LLMs}.
\newblock \emph{arXiv preprint arXiv:2410.06992}, 2024.
\newblock URL \url{https://arxiv.org/abs/2410.06992}.

\bibitem[Austin et~al.(2021)Austin, Odena, Nye, Bosma, Michalewski, Dohan,
  Jiang, Cai, Terry, Le, and Sutton]{austin2021mbpp}
Jacob Austin, Augustus Odena, Maxwell Nye, Maarten Bosma, Henryk Michalewski,
  David Dohan, Ellen Jiang, Carrie Cai, Michael Terry, Quoc Le, and Charles
  Sutton.
\newblock Program synthesis with large language models.
\newblock \emph{arXiv preprint arXiv:2108.07732}, 2021.
\newblock URL \url{https://arxiv.org/abs/2108.07732}.

\bibitem[Chan et~al.(2024)Chan, Chowdhury, Jaffe, Aung, Sherburn, Mays,
  Starace, Liu, Maksin, Patwardhan, Weng, and M{\k{a}}dry]{chan2024mlebench}
Jun~Shern Chan, Neil Chowdhury, Oliver Jaffe, James Aung, Dane Sherburn, Evan
  Mays, Giulio Starace, Kevin Liu, Leon Maksin, Tejal Patwardhan, Lilian Weng,
  and Aleksander M{\k{a}}dry.
\newblock {MLE-bench:} evaluating machine learning agents on machine learning
  engineering.
\newblock \emph{arXiv preprint arXiv:2410.07095}, 2024.
\newblock URL \url{https://arxiv.org/abs/2410.07095}.

\bibitem[Chen et~al.(2021)Chen, Tworek, Jun, Yuan, de~Oliveira~Pinto, Kaplan,
  Edwards, Burda, Joseph, Brockman, Ray, Puri, Krueger, Petrov, Khlaaf, Sastry,
  Mishkin, Chan, Gray, Ryder, Pavlov, Power, Kaiser, Bavarian, Winter, Tillet,
  Such, Cummings, Plappert, Chantzis, Barnes, Herbert-Voss, Guss, Nichol,
  Paino, Tezak, Tang, Babuschkin, Balaji, Jain, Saunders, Hesse, Carr, Leike,
  Achiam, Misra, Morikawa, Radford, Knight, Brundage, Murati, Mayer, Welinder,
  McGrew, Amodei, McCandlish, Sutskever, and Zaremba]{chen2021codex}
Mark Chen, Jerry Tworek, Heewoo Jun, Qiming Yuan, Henrique~Ponde
  de~Oliveira~Pinto, Jared Kaplan, Harri Edwards, Yuri Burda, Nicholas Joseph,
  Greg Brockman, Alex Ray, Raul Puri, Gretchen Krueger, Michael Petrov, Heidy
  Khlaaf, Girish Sastry, Pamela Mishkin, Brooke Chan, Scott Gray, Nick Ryder,
  Mikhail Pavlov, Alethea Power, Lukasz Kaiser, Mohammad Bavarian, Clemens
  Winter, Philippe Tillet, Felipe~Petroski Such, Dave Cummings, Matthias
  Plappert, Fotios Chantzis, Elizabeth Barnes, Ariel Herbert-Voss,
  William~Hebgen Guss, Alex Nichol, Alex Paino, Nikolas Tezak, Jie Tang, Igor
  Babuschkin, Suchir Balaji, Shantanu Jain, William Saunders, Christopher
  Hesse, Andrew~N. Carr, Jan Leike, Josh Achiam, Vedant Misra, Evan Morikawa,
  Alec Radford, Matthew Knight, Miles Brundage, Mira Murati, Katie Mayer, Peter
  Welinder, Bob McGrew, Dario Amodei, Sam McCandlish, Ilya Sutskever, and
  Wojciech Zaremba.
\newblock Evaluating large language models trained on code.
\newblock \emph{arXiv preprint arXiv:2107.03374}, 2021.
\newblock URL \url{https://arxiv.org/abs/2107.03374}.

\bibitem[Chen et~al.(2024)Chen, Chen, Ning, Zhang, Wang, Yu, Li, Liao, Wei, Lu,
  Dey, Xue, Baker, Burns, Adu-Ampratwum, Huang, Ning, Gao, Su, and
  Sun]{chen2024scienceagentbench}
Ziru Chen, Shijie Chen, Yuting Ning, Qianheng Zhang, Boshi Wang, Botao Yu,
  Yifei Li, Zeyi Liao, Chen Wei, Zitong Lu, Vishal Dey, Mingyi Xue, Frazier~N.
  Baker, Benjamin Burns, Daniel Adu-Ampratwum, Xuhui Huang, Xia Ning, Song Gao,
  Yu~Su, and Huan Sun.
\newblock {ScienceAgentBench:} toward rigorous assessment of language agents
  for {Data-Driven} scientific discovery.
\newblock \emph{arXiv preprint arXiv:2410.05080}, 2024.
\newblock URL \url{https://arxiv.org/abs/2410.05080}.

\bibitem[Dai et~al.(2024)Dai, Lu, Feng, Zeng, Ruan, Cheng, Huang, Tan, and
  Guo]{dai2024mhpp}
Jianbo Dai, Jianqiao Lu, Yunlong Feng, Guangtao Zeng, Rongju Ruan, Ming Cheng,
  Dong Huang, Haochen Tan, and Zhijiang Guo.
\newblock {MHPP:} exploring the capabilities and limitations of language models
  beyond basic code generation.
\newblock \emph{arXiv preprint arXiv:2405.11430}, 2024.
\newblock URL \url{https://arxiv.org/abs/2405.11430}.

\bibitem[Du et~al.(2023)Du, Liu, Wang, Wang, Liu, Chen, Feng, Sha, Peng, and
  Lou]{du2023classeval}
Xueying Du, Mingwei Liu, Kaixin Wang, Hanlin Wang, Junwei Liu, Yixuan Chen,
  Jiayi Feng, Chaofeng Sha, Xin Peng, and Yiling Lou.
\newblock {ClassEval:} {A} {Manually-Crafted} benchmark for evaluating {LLMs}
  on class-level code generation.
\newblock \emph{arXiv preprint arXiv:2308.01861}, 2023.
\newblock URL \url{https://arxiv.org/abs/2308.01861}.

\bibitem[Fakih et~al.(2024)Fakih, Dharmaji, Moghaddas, Araya, Ogundare, and
  Faruque]{fakih2024llm4plc}
Mohamad Fakih, Rahul Dharmaji, Yasamin Moghaddas, Gustavo~Quiros Araya,
  Oluwatosin Ogundare, and Mohammad Abdullah~Al Faruque.
\newblock {LLM4PLC:} harnessing large language models for verifiable
  programming of {PLCs} in industrial control systems.
\newblock \emph{arXiv preprint arXiv:2401.05443}, 2024.
\newblock URL \url{https://arxiv.org/abs/2401.05443}.

\bibitem[Gu et~al.(2024)Gu, Jiang, Shi, Tan, Zhai, Xu, Li, Shen, Ma, Liu, Wang,
  Zhang, Wang, Gao, Ni, and Guo]{gu2024judgesurvey}
Jiawei Gu, Xuhui Jiang, Zhichao Shi, Hexiang Tan, Xuehao Zhai, Chengjin Xu, Wei
  Li, Yinghan Shen, Shengjie Ma, Honghao Liu, Saizhuo Wang, Kun Zhang, Yuanzhuo
  Wang, Wen Gao, Lionel Ni, and Jian Guo.
\newblock {A} survey on {LLM-as-a-Judge}.
\newblock \emph{arXiv preprint arXiv:2411.15594}, 2024.
\newblock URL \url{https://arxiv.org/abs/2411.15594}.

\bibitem[Helff et~al.(2026)Helff, Delfosse, Steinmann, Härle, Shindo,
  Schramowski, Stammer, Kersting, and Friedrich]{helff2026gaming}
Lukas Helff, Quentin Delfosse, David Steinmann, Ruben Härle, Hikaru Shindo,
  Patrick Schramowski, Wolfgang Stammer, Kristian Kersting, and Felix
  Friedrich.
\newblock {LLMs} gaming verifiers: {RLVR} can lead to reward hacking.
\newblock \emph{arXiv preprint arXiv:2604.15149}, 2026.
\newblock URL \url{https://arxiv.org/abs/2604.15149}.

\bibitem[Jain et~al.(2024)Jain, Han, Gu, Li, Yan, Zhang, Wang, Solar-Lezama,
  Sen, and Stoica]{jain2024livecodebench}
Naman Jain, King Han, Alex Gu, Wen-Ding Li, Fanjia Yan, Tianjun Zhang, Sida
  Wang, Armando Solar-Lezama, Koushik Sen, and Ion Stoica.
\newblock {LiveCodeBench:} holistic and contamination free evaluation of large
  language models for code.
\newblock \emph{arXiv preprint arXiv:2403.07974}, 2024.
\newblock URL \url{https://arxiv.org/abs/2403.07974}.

\bibitem[Jimenez et~al.(2023)Jimenez, Yang, Wettig, Yao, Pei, Press, and
  Narasimhan]{jimenez2024swebench}
Carlos~E. Jimenez, John Yang, Alexander Wettig, Shunyu Yao, Kexin Pei, Ofir
  Press, and Karthik Narasimhan.
\newblock {SWE-bench:} can language models resolve {Real-World} {GitHub}
  issues?
\newblock \emph{arXiv preprint arXiv:2310.06770}, 2023.
\newblock URL \url{https://arxiv.org/abs/2310.06770}.

\bibitem[Jin et~al.(2025)Jin, Huang, Li, Cheng, Zhao, Zheng, Zhu, Xing, Dou,
  Zhang, Du, Guo, and Hu]{jin2025realbench}
Pengwei Jin, Di~Huang, Chongxiao Li, Shuyao Cheng, Yang Zhao, Xinyao Zheng,
  Jiaguo Zhu, Shuyi Xing, Bohan Dou, Rui Zhang, Zidong Du, Qi~Guo, and Xing Hu.
\newblock {RealBench:} benchmarking verilog generation models with {Real-World}
  {IP} designs.
\newblock \emph{arXiv preprint arXiv:2507.16200}, 2025.
\newblock URL \url{https://arxiv.org/abs/2507.16200}.

\bibitem[Liang et~al.(2025)Liang, Garg, and Moghaddam]{liang2025illusion}
Shanchao Liang, Spandan Garg, and Roshanak~Zilouchian Moghaddam.
\newblock The {SWE-Bench} illusion: When {State-of-the-Art} {LLMs} remember
  instead of reason.
\newblock \emph{arXiv preprint arXiv:2506.12286}, 2025.
\newblock URL \url{https://arxiv.org/abs/2506.12286}.

\bibitem[Liu et~al.(2023{\natexlab{a}})Liu, Xia, Wang, and
  Zhang]{liu2023evalplus}
Jiawei Liu, Chunqiu~Steven Xia, Yuyao Wang, and Lingming Zhang.
\newblock Is your code generated by {ChatGPT} really correct? rigorous
  evaluation of large language models for code generation.
\newblock \emph{arXiv preprint arXiv:2305.01210}, 2023{\natexlab{a}}.
\newblock URL \url{https://arxiv.org/abs/2305.01210}.

\bibitem[Liu et~al.(2023{\natexlab{b}})Liu, Pinckney, Khailany, and
  Ren]{liu2023verilogeval}
Mingjie Liu, Nathaniel Pinckney, Brucek Khailany, and Haoxing Ren.
\newblock {VerilogEval:} evaluating large language models for verilog code
  generation.
\newblock \emph{arXiv preprint arXiv:2309.07544}, 2023{\natexlab{b}}.
\newblock URL \url{https://arxiv.org/abs/2309.07544}.

\bibitem[Liu et~al.(2026{\natexlab{a}})Liu, Dai, Xuan, Chen, Zeng, Yang, Fang,
  Yang, Cui, Dong, Du, Zitnik, Zou, and Tang]{liu2026towards}
Tianyu Liu, Dingyuan Dai, Weihao Xuan, Jialin Chen, Qingcheng Zeng, Rui Yang,
  Ada Fang, Zhen Yang, Peng Cui, Yinpeng Dong, Yuanqi Du, Marinka Zitnik, James
  Zou, and Jie Tang.
\newblock Towards an agentic era of ai.
\newblock \emph{SSRN Electronic Journal}, 2026{\natexlab{a}}.
\newblock \doi{10.2139/ssrn.7245280}.

\bibitem[Liu et~al.(2026{\natexlab{b}})Liu, Wang, Panescu, Chen, Long, Wei,
  Jing, Zeng, Chen, Jiang, et~al.]{liu2026benchmarking}
Tianyu Liu, Allen~Xin Wang, Antonia Panescu, Lisa~Xinyi Chen, Wenxin Long,
  Xinyu Wei, Yueqian Jing, Ziyao Zeng, Jihang Chen, Sihan Jiang, et~al.
\newblock Benchmarking ai agents for addressing scientific challenges across
  scales.
\newblock \emph{arXiv preprint arXiv:2606.12736}, 2026{\natexlab{b}}.

\bibitem[Liu et~al.(2024)Liu, Zeng, Wang, Peng, Wang, Liu, Liu, and
  Wang]{liu2024agents4plc}
Zihan Liu, Ruinan Zeng, Dongxia Wang, Gengyun Peng, Jingyi Wang, Qiang Liu,
  Peiyu Liu, and Wenhai Wang.
\newblock {Agents4PLC:} automating closed-loop {PLC} code generation and
  verification in industrial control systems using {LLM-based} agents.
\newblock \emph{arXiv preprint arXiv:2410.14209}, 2024.
\newblock URL \url{https://arxiv.org/abs/2410.14209}.

\bibitem[Lu et~al.(2023)Lu, Liu, Zhang, and Xie]{lu2024rtllm}
Yao Lu, Shang Liu, Qijun Zhang, and Zhiyao Xie.
\newblock {RTLLM:} an {Open-Source} benchmark for design {RTL} generation with
  large language model.
\newblock \emph{arXiv preprint arXiv:2308.05345}, 2023.
\newblock URL \url{https://arxiv.org/abs/2308.05345}.

\bibitem[Nishimura-Gasparian et~al.(2026)Nishimura-Gasparian, McCarthy, and
  Lindner]{nishimura2026specgaming}
Kei Nishimura-Gasparian, Robert McCarthy, and David Lindner.
\newblock Towards understanding specification gaming in reasoning models.
\newblock \emph{arXiv preprint arXiv:2605.02269}, 2026.
\newblock URL \url{https://arxiv.org/abs/2605.02269}.

\bibitem[Pan et~al.(2024)Pan, Wang, Neubig, Jaitly, Ji, Suhr, and
  Zhang]{pan2024swegym}
Jiayi Pan, Xingyao Wang, Graham Neubig, Navdeep Jaitly, Heng Ji, Alane Suhr,
  and Yizhe Zhang.
\newblock Training software engineering agents and verifiers with {SWE-Gym}.
\newblock \emph{arXiv preprint arXiv:2412.21139}, 2024.
\newblock URL \url{https://arxiv.org/abs/2412.21139}.

\bibitem[Ray(2026)]{ray2026fuzzing}
Jaideep Ray.
\newblock Before the model learns the {Bug:Fuzzing} {RLVR} verifiers.
\newblock \emph{arXiv preprint arXiv:2606.01066}, 2026.
\newblock URL \url{https://arxiv.org/abs/2606.01066}.

\bibitem[Sainz et~al.(2023)Sainz, Campos, García-Ferrero, Etxaniz, de~Lacalle,
  and Agirre]{sainz2023contamination}
Oscar Sainz, Jon~Ander Campos, Iker García-Ferrero, Julen Etxaniz, Oier~Lopez
  de~Lacalle, and Eneko Agirre.
\newblock {NLP} evaluation in trouble: On the need to measure {LLM} data
  contamination for each benchmark.
\newblock \emph{arXiv preprint arXiv:2310.18018}, 2023.
\newblock URL \url{https://arxiv.org/abs/2310.18018}.

\bibitem[Sharifloo et~al.(2025)Sharifloo, Heydari, Kazerooni, Maninger, and
  Mezini]{sharifloo2025struggle}
Amir~Molzam Sharifloo, Maedeh Heydari, Parsa Kazerooni, Daniel Maninger, and
  Mira Mezini.
\newblock Where do {LLMs} still struggle? an {In-Depth} analysis of code
  generation benchmarks.
\newblock \emph{arXiv preprint arXiv:2511.04355}, 2025.
\newblock URL \url{https://arxiv.org/abs/2511.04355}.

\bibitem[Siegel et~al.(2024)Siegel, Kapoor, Nadgir, Stroebl, and
  Narayanan]{siegel2024corebench}
Zachary~S. Siegel, Sayash Kapoor, Nitya Nadgir, Benedikt Stroebl, and Arvind
  Narayanan.
\newblock {CORE-Bench:} fostering the credibility of published research through
  a computational reproducibility agent benchmark.
\newblock \emph{arXiv preprint arXiv:2409.11363}, 2024.
\newblock URL \url{https://arxiv.org/abs/2409.11363}.

\bibitem[Stroebl et~al.(2024)Stroebl, Kapoor, and
  Narayanan]{stroebl2024resampling}
Benedikt Stroebl, Sayash Kapoor, and Arvind Narayanan.
\newblock The limits of inference scaling through resampling.
\newblock \emph{arXiv preprint arXiv:2411.17501}, 2024.
\newblock URL \url{https://arxiv.org/abs/2411.17501}.

\bibitem[Vangala et~al.(2025)Vangala, Adibifar, Gehani, and
  Malik]{vangala2025reproducible}
Bhanu~Prakash Vangala, Ali Adibifar, Ashish Gehani, and Tanu Malik.
\newblock {AI-Generated} code is not reproducible (yet): An empirical study of
  dependency gaps in {LLM-Based} coding agents.
\newblock \emph{arXiv preprint arXiv:2512.22387}, 2025.
\newblock URL \url{https://arxiv.org/abs/2512.22387}.

\bibitem[Yamauchi et~al.(2025)Yamauchi, Yano, and
  Oyamada]{yamauchi2025judgedesign}
Yusuke Yamauchi, Taro Yano, and Masafumi Oyamada.
\newblock An empirical study of {LLM-as-a-Judge:} how design choices impact
  evaluation reliability.
\newblock \emph{arXiv preprint arXiv:2506.13639}, 2025.
\newblock URL \url{https://arxiv.org/abs/2506.13639}.

\bibitem[Yang et~al.(2024)Yang, Jimenez, Wettig, Lieret, Yao, Narasimhan, and
  Press]{yang2024sweagent}
John Yang, Carlos~E. Jimenez, Alexander Wettig, Kilian Lieret, Shunyu Yao,
  Karthik Narasimhan, and Ofir Press.
\newblock {SWE-agent:} {Agent-Computer} interfaces enable automated software
  engineering.
\newblock \emph{arXiv preprint arXiv:2405.15793}, 2024.
\newblock URL \url{https://arxiv.org/abs/2405.15793}.

\bibitem[Yang et~al.(2025)Yang, Lieret, Jimenez, Wettig, Khandpur, Zhang, Hui,
  Press, Schmidt, and Yang]{yang2025swesmith}
John Yang, Kilian Lieret, Carlos~E. Jimenez, Alexander Wettig, Kabir Khandpur,
  Yanzhe Zhang, Binyuan Hui, Ofir Press, Ludwig Schmidt, and Diyi Yang.
\newblock {SWE-smith:} scaling data for software engineering agents.
\newblock \emph{arXiv preprint arXiv:2504.21798}, 2025.
\newblock URL \url{https://arxiv.org/abs/2504.21798}.

\bibitem[Zhao et~al.(2026)Zhao, Srikanth, Wu, and Jiang]{zhao2026specbench}
Bingchen Zhao, Dhruv Srikanth, Yuxiang Wu, and Zhengyao Jiang.
\newblock {SpecBench:} measuring reward hacking in {Long-Horizon} coding
  agents.
\newblock \emph{arXiv preprint arXiv:2605.21384}, 2026.
\newblock URL \url{https://arxiv.org/abs/2605.21384}.

\bibitem[Zhao et~al.(2024{\natexlab{a}})Zhao, Huang, Lv, Cui, Sun, Mao, Zhang,
  Xin, Yin, Li, and Wei]{zhao2024mmlucf}
Qihao Zhao, Yangyu Huang, Tengchao Lv, Lei Cui, Qinzheng Sun, Shaoguang Mao,
  Xin Zhang, Ying Xin, Qiufeng Yin, Scarlett Li, and Furu Wei.
\newblock {MMLU-CF:} {A} contamination-free multi-task language understanding
  benchmark.
\newblock \emph{arXiv preprint arXiv:2412.15194}, 2024{\natexlab{a}}.
\newblock URL \url{https://arxiv.org/abs/2412.15194}.

\bibitem[Zhao et~al.(2024{\natexlab{b}})Zhao, Jiang, Lee, Chiu, Cardie, Gallé,
  and Rush]{zhao2024commit0}
Wenting Zhao, Nan Jiang, Celine Lee, Justin~T Chiu, Claire Cardie, Matthias
  Gallé, and Alexander~M Rush.
\newblock {Commit0:} library generation from scratch.
\newblock \emph{arXiv preprint arXiv:2412.01769}, 2024{\natexlab{b}}.
\newblock URL \url{https://arxiv.org/abs/2412.01769}.

\bibitem[Zheng et~al.(2023)Zheng, Chiang, Sheng, Zhuang, Wu, Zhuang, Lin, Li,
  Li, Xing, Zhang, Gonzalez, and Stoica]{zheng2023judging}
Lianmin Zheng, Wei-Lin Chiang, Ying Sheng, Siyuan Zhuang, Zhanghao Wu, Yonghao
  Zhuang, Zi~Lin, Zhuohan Li, Dacheng Li, Eric~P. Xing, Hao Zhang, Joseph~E.
  Gonzalez, and Ion Stoica.
\newblock Judging {LLM-as-a-Judge} with {MT-Bench} and chatbot arena.
\newblock \emph{arXiv preprint arXiv:2306.05685}, 2023.
\newblock URL \url{https://arxiv.org/abs/2306.05685}.

\bibitem[Zhou et~al.(2023)Zhou, Zhu, Chen, Chen, Zhao, Chen, Lin, Wen, and
  Han]{zhou2023cheater}
Kun Zhou, Yutao Zhu, Zhipeng Chen, Wentong Chen, Wayne~Xin Zhao, Xu~Chen,
  Yankai Lin, Ji-Rong Wen, and Jiawei Han.
\newblock Don't make your {LLM} an evaluation benchmark cheater.
\newblock \emph{arXiv preprint arXiv:2311.01964}, 2023.
\newblock URL \url{https://arxiv.org/abs/2311.01964}.

\bibitem[Zhuo et~al.(2024)Zhuo, Vu, Chim, Hu, Yu, Widyasari, Yusuf, Zhan, He,
  Paul, Brunner, Gong, Hoang, Zebaze, Hong, Li, Kaddour, Xu, Zhang, Yadav,
  Jain, Gu, Cheng, Liu, Liu, Wang, Hui, Muennighoff, Lo, Fried, Du, de~Vries,
  and Werra]{zhuo2024bigcodebench}
Terry~Yue Zhuo, Minh~Chien Vu, Jenny Chim, Han Hu, Wenhao Yu, Ratnadira
  Widyasari, Imam Nur~Bani Yusuf, Haolan Zhan, Junda He, Indraneil Paul, Simon
  Brunner, Chen Gong, Thong Hoang, Armel~Randy Zebaze, Xiaoheng Hong, Wen-Ding
  Li, Jean Kaddour, Ming Xu, Zhihan Zhang, Prateek Yadav, Naman Jain, Alex Gu,
  Zhoujun Cheng, Jiawei Liu, Qian Liu, Zijian Wang, Binyuan Hui, Niklas
  Muennighoff, David Lo, Daniel Fried, Xiaoning Du, Harm de~Vries, and
  Leandro~Von Werra.
\newblock {BigCodeBench:} benchmarking code generation with diverse function
  calls and complex instructions.
\newblock \emph{arXiv preprint arXiv:2406.15877}, 2024.
\newblock URL \url{https://arxiv.org/abs/2406.15877}.

\end{thebibliography}

\newpage 

\appendix

\section{Full results tables}\label{app:tables}

\begin{table}[ht]
\caption{Clean-room reconstruction of thirteen industrial software engines (the data behind
Figure~\ref{fig:results}a). Each cell is the hidden-verifier score of the best run under the
\emph{protocol} condition; scores are comparable \emph{within} a row only; \cmark{} denotes a pass.
GLM-5.3 flash cells use the runs described in Section~\ref{sec:threats}. Pass marks denote
task-level thresholds, not necessarily perfect scores. CFD scores measure interface compliance
and do not establish numerical-solver capability (Section~\ref{sec:verifier}); MES and PLM do
not distinguish the evaluated models.}
\label{tab:matrix}
\centering
\resizebox{\linewidth}{!}{%
\begin{tabular}{ll cccc @{\hspace{1em}} ccc @{\hspace{1em}} c}
\toprule
& & \multicolumn{4}{c}{Anthropic} & \multicolumn{3}{c}{OpenAI GPT-5.6} & Zhipu \\
\cmidrule(lr){3-6}\cmidrule(lr){7-9}\cmidrule(lr){10-10}
Engine & Commercial analog & Fable 5.1 & Opus 5 & Sonnet 5 & Haiku 4.5 & sol & luna & terra & GLM-5.3 flash \\
\midrule
Statistics engine & SAS, SPSS & \cmark\,\scriptsize 43/43 & \cmark\,\scriptsize 43/43 & \xmark\,\scriptsize 34/43 & \xmark\,\scriptsize 23/43 & \cmark\,\scriptsize 43/43 & \cmark\,\scriptsize 40/43 & \cmark\,\scriptsize 41/43 & \cmark\,\scriptsize 41/43 \\
Circuit simulator & HSPICE & \cmark\,\scriptsize 10/10 & \cmark\,\scriptsize 9/10 & \cmark\,\scriptsize 9/10 & \xmark\,\scriptsize 0/10 & \xmark\,\scriptsize 7/10 & \xmark\,\scriptsize 5/10 & \xmark\,\scriptsize 6/10 & \xmark\,\scriptsize 0/10 \\
PLC runtime & TIA Portal & \cmark\,\scriptsize 9/10 & \cmark\,\scriptsize 9/10 & \cmark\,\scriptsize 10/10 & \xmark\,\scriptsize 6/10 & \cmark\,\scriptsize 10/10 & \xmark\,\scriptsize 2/10 & \cmark\,\scriptsize 10/10 & \cmark\,\scriptsize 8/10 \\
FE solver & Nastran & \cmark\,\scriptsize 4/4 & \cmark\,\scriptsize 4/4 & \cmark\,\scriptsize 4/4 & \xmark\,\scriptsize 2/4 & \xmark\,\scriptsize 2/4 & \cmark\,\scriptsize 4/4 & \xmark\,\scriptsize 2/4 & \cmark\,\scriptsize 4/4 \\
2D CAD kernel & AutoCAD & \cmark\,\scriptsize 31/31 & \cmark\,\scriptsize 31/31 & \cmark\,\scriptsize 29/31 & \xmark\,\scriptsize 0/31 & \cmark\,\scriptsize 29/31 & \xmark\,\scriptsize 0/31 & \xmark\,\scriptsize 21/31 & \xmark\,\scriptsize 0/31 \\
Logic synthesizer & Design Compiler & \cmark\,\scriptsize 5/5 & \cmark\,\scriptsize 5/5 & \xmark\,\scriptsize 4/5 & \xmark\,\scriptsize 2/5 & \cmark\,\scriptsize 5/5 & \xmark\,\scriptsize 3/5 & \xmark\,\scriptsize 3/5 & \xmark\,\scriptsize 1/5 \\
3D solid modeler & Parasolid & \cmark\,\scriptsize 5/5 & \cmark\,\scriptsize 5/5 & \cmark\,\scriptsize 5/5 & \xmark\,\scriptsize 2/5 & \cmark\,\scriptsize 5/5 & \xmark\,\scriptsize 2/5 & \xmark\,\scriptsize 4/5 & \xmark\,\scriptsize 1/5 \\
ERP core & SAP S/4HANA & \cmark\,\scriptsize 5/5 & \cmark\,\scriptsize 5/5 & \xmark\,\scriptsize 4/5 & \xmark\,\scriptsize 2/5 & \cmark\,\scriptsize 5/5 & \cmark\,\scriptsize 5/5 & \cmark\,\scriptsize 5/5 & \xmark\,\scriptsize 4/5 \\
Place \& route & Innovus & \cmark\,\scriptsize 5/5 & \cmark\,\scriptsize 5/5 & \cmark\,\scriptsize 5/5 & \xmark\,\scriptsize 2/5 & \cmark\,\scriptsize 5/5 & \cmark\,\scriptsize 5/5 & \xmark\,\scriptsize 2/5 & \cmark\,\scriptsize 5/5 \\
CFD / CAE & Fluent & \cmark\,\scriptsize 6/6 & \cmark\,\scriptsize 6/6 & \cmark\,\scriptsize 6/6 & \xmark\,\scriptsize 2/6 & \xmark\,\scriptsize 4/6 & \xmark\,\scriptsize 4/6 & \xmark\,\scriptsize 4/6 & \xmark\,\scriptsize 4/6 \\
SCADA/DCS runtime & WinCC & \cmark\,\scriptsize 6/6 & \cmark\,\scriptsize 6/6 & \cmark\,\scriptsize 6/6 & \xmark\,\scriptsize 2/6 & \xmark\,\scriptsize 2/6 & \xmark\,\scriptsize 5/6 & \xmark\,\scriptsize 2/6 & \cmark\,\scriptsize 6/6 \\
MES & SAP ME & \cmark\,\scriptsize 6/6 & \cmark\,\scriptsize 6/6 & \cmark\,\scriptsize 6/6 & \cmark\,\scriptsize 6/6 & \cmark\,\scriptsize 6/6 & \cmark\,\scriptsize 6/6 & \cmark\,\scriptsize 6/6 & \cmark\,\scriptsize 6/6 \\
PLM / BOM & Teamcenter & \cmark\,\scriptsize 7/7 & \cmark\,\scriptsize 7/7 & \cmark\,\scriptsize 7/7 & \cmark\,\scriptsize 7/7 & \cmark\,\scriptsize 7/7 & \cmark\,\scriptsize 7/7 & \cmark\,\scriptsize 7/7 & \cmark\,\scriptsize 7/7 \\
\midrule
\textbf{Engines passed} & & \textbf{13/13} & \textbf{13/13} & \textbf{10/13} & \textbf{2/13} & \textbf{9/13} & \textbf{6/13} & \textbf{5/13} & \textbf{7/13} \\
\bottomrule\end{tabular}}\end{table}

\begin{table}[ht]
\caption{Per-model totals on the reconstruction suite. ``Engines passed'' is taken from
Table~\ref{tab:matrix}; runs, turns and cost are over all valid runs of that model.}
\label{tab:models}
\centering\small
\begin{tabular}{lrrrr}
\toprule
Model & Engines passed & Valid runs & Mean turns & Cost (USD) \\
\midrule
Fable 5.1 & 13/13 & 13 & 20.8 & 95 \\
Opus 5 & 13/13 & 9 & 74.1 & 76 \\
Sonnet 5 & 10/13 & 13 & 60.1 & 35 \\
Haiku 4.5 & 2/13 & 15 & 80.3 & 15 \\
sol & 9/13 & 13 & 30.2 & n/a$^{\dagger}$ \\
luna & 6/13 & 13 & 35.3 & n/a$^{\dagger}$ \\
terra & 5/13 & 13 & 31.0 & n/a$^{\dagger}$ \\
GLM-5.3 flash & 7/13 & 13$^{\ddagger}$ & 83.1$^{\ddagger}$ & n/a$^{\dagger}$ \\
\bottomrule\end{tabular}
\\[2pt]\footnotesize $^{\dagger}$ Run through a local protocol adapter, so the harness prices them at the
launch model's rate; that figure is an artifact and is withheld. Token counts are genuine.
$^{\ddagger}$ The thirteen runs used in the matrix; four hit the wall-clock limit and record no turn
count, so the mean is over the other nine.
\end{table}

\begin{table}[t]
\caption{Reasoning-effort ablation (Opus 5, five levels). Scores remain unchanged. Cost, turns, and wall time
increase monotonically for ERP but fluctuate at intermediate settings for CAD. Both tasks are
saturated at all five settings, limiting conclusions about improvements in accuracy (Appendix~\ref{app:effort}).}
\label{tab:effort}
\centering\small
\begin{tabular}{llrrrr}
\toprule
Task & Effort & Score & Turns & Wall (s) & Cost (USD) \\
\midrule
ERP core & low & 5/5 & 9 & 204 & 1.00 \\
 & medium & 5/5 & 18 & 411 & 1.85 \\
 & high & 5/5 & 20 & 672 & 2.71 \\
 & xhigh & 5/5 & 30 & 1070 & 4.37 \\
 & max & 5/5 & 31 & 1464 & 4.79 \\
\midrule
2D CAD kernel & low & 31/31 & 65 & 1426 & 6.95 \\
 & medium & 31/31 & 64 & 2242 & 8.09 \\
 & high & 31/31 & 46 & 1914 & 7.09 \\
 & xhigh & 31/31 & 101 & 3117 & 13.36 \\
 & max & 31/31 & 121 & 4588 & 15.93 \\
\bottomrule\end{tabular}\end{table}

\section{Reasoning-effort ablation}\label{app:effort}

We varied the agent's reasoning-effort setting across its five levels on two tasks, after first
confirming on the wire that the setting reaches the API and that all five levels produce distinct
requests. Table~\ref{tab:effort} shows unchanged scores at all five levels. Cost, turns, and wall time
increase monotonically for ERP, but not for CAD. On CAD, the highest setting costs $2.3\times$
as much as the lowest and uses $2.2\times$ as many output tokens, despite intermediate fluctuations.

We do not read this as ``effort does not matter''. Post-hoc analysis shows the model already
achieves full sub-scores on both tasks at the default setting, which leaves no room for higher scores at that setting. The unchanged scores therefore
provide limited evidence about the value of additional reasoning effort. The second arm was chosen \emph{because} it appeared
to have headroom ($29/31$), and the two missing operations turned out to be two of our own verifier
defects; once fixed, all five levels score $31/31$. This is the same rule as our difficulty probe,
pointed at the other end of the ability range: \textbf{saturated tasks have limited power to reveal improvements in accuracy.}
They can still reveal differences in resource use. Here, the ablation also had diagnostic value:
identical failures across five settings prompted investigation of the same two verifier defects.

\end{document}